\documentstyle[12pt]{article}
\begin{document}
\def\theequation{\arabic{section}.\arabic{equation}}
\newcommand{\be}{\begin{equation}}
\newcommand{\ee}{\end{equation}}
\begin{titlepage}
\setcounter{page}{1}
\title{On the total energy of open Friedmann-Robertson-Walker universes}
\author{V. Faraoni$^{1,2}$ and F.I. Cooperstock$^3$ \\ \\
{\small \it $^1$Physics Department, University of Northern British
Columbia} \\ 
{\small \it 3333 University Way, Prince George, B.C., Canada, V2N~4Z9}\\
{\small \it electronic mail: vfaraoni@unbc.ca}\\
{\small \it $^2$Department of Physics and Astronomy, University of
Victoria}\\ 
{\small \it P.O. Box 3055, Victoria, B.C. Canada V8W 3P6}\\
{\small \it $^3$Department of Physics and Astronomy, University of Victoria}\\
{\small \it P.O. Box 3055, Victoria, B.C. Canada V8W 3P6}\\
{\small \it electronic mail: cooperstock@phys.uvic.ca}
}
\date{}
\maketitle   
\end{titlepage}   \clearpage \setcounter{page}{2}

\begin{center} {\bf Abstract} \end{center}

The idea that the universe has zero total energy when one includes the 
contribution from the gravitational field is reconsidered.  A Hamiltonian
is proposed as an energy for the exact equations of FRW
cosmology: it is then shown that this energy is constant. 
Thus open and critically  open FRW universes have the energy 
of their asymptotic state of infinite dilution, which is Minkowski
space with zero energy. It is then shown that de Sitter space, the
inflationary attractor, also has zero energy, 
and the argument is generalized to Bianchi models converging to this 
attractor.

\clearpage

\section{Introduction}

\setcounter{equation}{0}

Over the years, there have been many suggestions and assertions in the 
literature that the total
energy of the universe, that which includes not only the material energy but 
also the contribution from the gravitational field, should be zero. The
subject
is non-trivial partially due to the fact that in general relativity there
is no universally accepted
prescription for the localization of energy including the contribution from 
gravity. In the case 
of a cosmological
metric, a rationalized energy localization would provide an expression to 
integrate over three-dimensional
spatial sections, yielding the total energy of the universe. In
principle, this could be  infinite for an open universe and this would likely
be the instinctive response of most physicists.

In this paper we do not address the energy localization problem but rather
we focus upon the {\em total} energy of the
universe. The idea that the universe should have zero total energy dates back
many years (a topological argument due to P. Bergmann is quoted in
Tryon 1973) and returned to the fore with the works of Albrow
(1973) and Tryon (1973)
suggesting that the universe originated as a quantum fluctuation of the vacuum. 
The idea was
developed in Guth's theory of inflation (Guth 1981) that has had a 
major impact on modern cosmology. Further, an approach to quantum
cosmology proposed
that the universe was born as the result of quantum tunneling from nothing
(Vilenkin 1983).

The question as to whether or not the total energy of the universe is zero 
was re-opened by 
Rosen (1994), Cooperstock (1994),
Cooperstock \& Israelit (1995), Johri {\em et al.} (1995), 
Banerjee \& Sen (1997), Radinschi (1999), and Xulu (2000),
who studied both Friedmann-Robertson-Walker (hereafter referred to as FRW)
universes and anisotropic Bianchi models. The approaches followed by these 
authors
are varied. Apart from Cooperstock (1995) and Cooperstock \& Israelit (1995), a
common feature 
is the use of pseudotensors which brings these
approaches into question. Pseudotensors have been shown to be useful tools 
in the study of bounded systems which are asymptotically Minkowskian 
(in Cartesian coordinates).
However, for cosmological metrics, we are concerned with systems that
are infinite or do not possess an asymptotically flat exterior. In
spite of this, it is note-worthy that
these authors invariably  conclude that
the total energy of open and closed FRW universes, as well as of Bianchi 
models, vanishes.

In this paper, open FRW cosmologies are discussed and it is shown that for 
many equations of state (both time-independent and time-dependent), the total 
energy is constant and, one could argue, has the value zero. The essential
elements
that are employed to make 
this deduction are the global conservation of energy and the vanishing of the 
energy of flat spacetime. Energy conservation, including the contribution
from gravity, is built into the Einstein
field equations by virtue of the vanishing of the covariant divergence of 
the energy-momentum tensor. That flat spacetime represents the absence of
energy can
be seen as follows. The mass parameter $m$ of the Schwarzschild metric 
\footnote{We use $\kappa \equiv 8\pi G$, where $G$ is
Newton's constant; the metric signature  is --~+~+~+, the speed of light and
Planck's constant assume the
value unity.  The components of the Ricci tensor are given in terms of the
Christoffel symbols
$\Gamma_{\alpha\beta}^{\delta}$ 
by $R_{\mu\rho}=
\Gamma^{\nu}_{\mu\rho ,\nu}-\Gamma^{\nu}_{\nu\rho ,\mu}+
\Gamma^{\alpha}_{\mu\rho}\Gamma^{\nu}_{\alpha\nu}-
\Gamma^{\alpha}_{\nu\rho}\Gamma^{\nu}_{\alpha\mu} $, and an overdot
denotes differentiation with
respect to the comoving cosmic time $t$.}
\be
ds^2=-\left( 1-\frac{2m}{r} \right) dt^2+
\left( 1-\frac{2m}{r} \right)^{-1}dr^2 +r^2 \left( d\theta^2+\sin^2 \theta\,
d\varphi^2 \right)
\ee
yields the total energy including the contribution from gravity.
When $m$ is set
to zero, the metric becomes flat and hence flat spacetime has zero
energy.

We assume that the energy 
density of the material content
of the universe $\rho$ is non-negative and we consider an open or 
critically open (the case favoured by inflationary theories and by
current cosmological data (Liddle \& Lyth 2000))  
FRW universe 
described by the line element 
\be  \label{1}
ds^2=-dt^2+a^2 (t) \left[ \, \frac{dr^2}{1-Kr^2}+r^2 \left( d\theta^2+\sin^2
\theta\, d\varphi^2 \right) \right] 
\ee
in comoving coordinates $\left( t,r,\theta,\varphi \right)$, where the
curvature index $K$ can assume the values $- 1$ or 0.

\section{An explicit energy definition}
\setcounter{equation}{0}

If the total energy of the universe is constant for a certain form of its
material content, the same should be true when  the same
cosmological metric is generated by a different form of matter.

Following the dynamical systems approach to cosmology  
(Stabell \& Refsdal 1966, Madsen
\& Ellis 1988, Madsen {\em et al.} 1992, Wainwright \& Ellis 1997;
Amendola {\em  et al.} 1990, Foster 1988, Gunzig {\em et al.} 2000,
Rocha-Filho {\em et al.} 2000, Gunzig {\em et al.} 2001{\em a}, 2001{\em
b}), we consider a $K=0$ FRW universe with 
a scalar field $\phi(t)$ as the only source of gravity; $\phi$ is allowed
to couple nonminimally to the Ricci curvature as described by the action
\be\label{action}
S=\int d^4x \, \sqrt{-g}\,\left[ \frac{1}{2} \left( \frac{1}{\kappa}
-\xi\, \phi^2 \right) R -\, \frac{1}{2} \,
\nabla^{\mu} \phi \nabla_{\mu} \phi -V( \phi)  \right]  \;,
\ee
where $\xi$ is a dimensionless coupling constant and $V( \phi) $ is the
scalar
field potential. $H \equiv \dot{a}/a $ and $\phi$ are
chosen as dynamical variables and the relevant dynamical system is 
\be  \label{fe1}
6\left[ 1 -\xi \left( 1- 6\xi \right) \kappa \phi^2
\right] \left( \dot{H} +2H^2 \right) 
-\kappa \left( 6\xi -1 \right) \dot{\phi}^2   
- 4 \kappa V  + 6\kappa \xi \phi \, \frac{dV}{d\phi} = 0 \; ,
\ee
\begin{equation}  \label{fe2}
\frac{\kappa}{2}\,\dot{\phi}^2 + 6\xi\kappa H\phi\dot{\phi}
- 3H^2 \left( 1-\kappa \xi \phi^2 \right) + \kappa  V =0 \, ,
\end{equation}
\be  \label{KG}
\ddot{\phi}+3H\dot{\phi}+\xi R \phi +\frac{dV}{d\phi} =0 \; .
\ee
Let us restrict now to the case of a conformally coupled scalar, 
obtained by setting $\xi=1/6$ in the action (\ref{action}). 
This value of the coupling constant is dictated by physical requirements
such as the Einstein equivalence principle (Sonego and
Faraoni 1993, Grib and Poberii 1995, 
Grib and Rodrigues 1996), 
renormalizability (Callan, Coleman and Jackiw
1970), or the requirement  that $\xi$, which is a running coupling at high
energies, sit in a  
stable infrared fixed point of the renormalization group 
(Parker and Toms 1985, Buchbinder, Odintsov and Shapiro 1992). We allow
the scalar 
field to acquire a mass and include in the picture a quartic
self-interaction plus
(motivated by the role of de Sitter attractors in phase
space) the possibility 
of a cosmological constant, as described by the potential
\be
V( \phi) = \frac{m^2\, \phi^2}{2}+\lambda \phi^4 +V_0 \;.
\ee 
By introducing the new variables (Rocha Filho {\em et al.} 2000)
\be
\psi \equiv \sqrt{ \frac{\kappa}{6}} \,\, a \;,
\ee
\be 
\varphi \equiv a\, \phi \;,
\ee
and the conformal time $\eta $ defined by $dt \equiv a \,d\eta$, the field
equations  (\ref{fe1}) and (\ref{KG}) become
\be \label{newfe1}
\psi'' -\frac{\kappa m^2}{6} \, \varphi^2 \psi -4 V_0 \psi^3=0 \;,
\ee
\be  \label{newfe3}
\varphi'' +\frac{6m^2}{\kappa} \, \varphi \, \psi^2 +4\lambda \varphi^3 =0
\;,
\ee
where a prime denotes differentiation with respect to the 
conformal time $\eta$.  Eqs.~(\ref{newfe1}) and
(\ref{newfe3}) can be derived from the Lagrangian
\be
\L=\frac{1}{2} \left( \varphi ' \right)^2 
-\frac{18}{\kappa^2} \, \left( \psi '\right)^2 
-\frac{3m^2}{\kappa}\, \varphi^2\psi^2 -\lambda \varphi^4 
-\frac{36}{\kappa^2} \, V_0 \psi^4 \; ,
\ee
while the equation
\be  \label{newfe2}
\frac{1}{2}\, \left( \varphi ' \right)^2 -\frac{18}{\kappa^2}\, \left( \psi' \right)^2 
+\frac{3m^2}{\kappa} \, \varphi^2 \psi^2 
+\lambda \varphi^4 +\frac{36}{\kappa^2} \, V_0 \psi^4=\mbox{const.}
\ee
is obtained by manipulating  eqs.~(\ref{newfe1}) and (\ref{newfe3})
and integrating once.

The momenta canonically conjugated to the variables $\varphi$ and $\psi$
are 
$p_{\varphi} \equiv \partial L/\partial ( \varphi')=\varphi'$ and 
$p_{\psi} \equiv \partial L/\partial (\psi' )=-36\psi'/\kappa^2 $. It is
straightforward to check 
that, by using the associated Hamiltonian
\be   \label{Hamiltonian}
E= \frac{1}{2} \, \left( \varphi' \right)^2 
-\frac{18}{\kappa^2}\, \left( \psi' \right)^2
+\frac{3m^2}{\kappa^2} \, \varphi^2\psi^2 
+\lambda \varphi^4
+\frac{36}{\kappa^2}\, V_0 \psi^4 \;,
\ee
the Hamilton equations 
\be 
p_{\varphi}'=-\, \frac{\partial E}{\partial \varphi} \; , \;\;\;\;\;\;
p_{\psi}'=-\, \frac{\partial E}{\partial \psi} \;,
\ee
reproduce the field equations (\ref{newfe1}) 
and (\ref{newfe3}). Furthermore, eq.~(\ref{newfe2}) 
states that 
\be
E=\mbox{constant} \;.
\ee
As customary in classical mechanics and in quantum physics, one identifies the 
Hamiltonian with the energy of 
the system, thus obtaining a total 
energy of the universe (\ref{Hamiltonian}) that is conserved. 
Note that de Sitter space is among the possible solutions of the field
equations 
(\ref{newfe1}), (\ref{newfe3}), and (\ref{KG}), and therefore it has
constant energy as
well.
This concludes the first step of the deduction of zero energy for FRW
universes, while the next section presents the second step.

\section{The limit to Minkowski spacetime}

\setcounter{equation}{0}

Open or critically open FRW cosmologies have
Minkowski space as their asymptotic state. In fact, the
dynamical system (\ref{fe1})-(\ref{KG}) has the
Minkowski space $\left( H, \phi \right)=\left( 0,0 \right)$ as a fixed
point, with attractive behaviour in the half-plane  $H>0$ and repulsive behaviour
for $H<0$ (Gunzig {\em et al.} 2001{\em a}). It was speculated that, when
$V_0=0$, the universe could have emerged  from points
arbitrarily
close to the Minkowski fixed point, a classical analog of the proposals of
Albrow (1973), Tryon (1973) and of Prigogine {\em et al.} (1988, 1989).
The same Minkowski fixed point is an
attractor at large times, in a
certain basin, for solutions expanding into infinite
dilution and asymptotically approaching $\left( H,
\phi \right)=\left( 0,0 \right)$.

Consider first an open ($K=-1$) or critically open ($K=0$) universe; its 
asymptotic state at large times is one of infinite dilution. In fact,
the
conservation equation (Liddle \& Lyth 2000, Kolb \& Turner 1994, Weinberg 1972, 
Landau \& Lifshitz 1989) 
\be \label{conservation}
\dot{\rho}+3H\left( P+\rho \right)=0
\ee
yields $\dot{\rho} <0 $ when  $P \geq -\rho /3$ and the
universe expands. Since $\rho$ is bounded from below by zero and is
monotonically decreasing, one has  $\rho(t)
\rightarrow 0$ at large times: the asymptotic state is one of infinite dilution
corresponding to  Minkowski space. The Hubble radius is given by the
Friedmann equation
\be \label{3}
H^2= \frac{\kappa \, \rho}{3} -\, \frac{K}{a^2}  
\ee 
and is
\be
H^{-1}=\left( \frac{\kappa \rho}{3}-\, \frac{K}{a^2} \right)^{-1/2} \; ;
\ee
it is the only characteristic geometrical scale and diverges as
$t\rightarrow +\infty$.  This property,
familiar in
the particular case of a matter- or radiation-dominated era in a $K=0$ FRW
universe, is quite general.

In principle the situation could be different when $P< -\rho/3$ with the 
universe
undergoing accelerated expansion, $\ddot{a}>0$. Power-law solutions with
scale factor $
a(t)=a_0 \, t^p $ and $ p=2/ \left( 3\gamma \right)$ 
are obtained for $K=0$ by imposing the equation of state $P=\left(
\gamma -1 \right) \rho$. These solutions exhibit divergent Hubble radius
$H^{-1}=t/p$, have Minkowski
spacetime as their limit at large times if $\gamma>0$, and are attractors
in phase
space.
The property of future convergence to Minkowski spacetime is shared by all
solutions satisfying $P>-\rho$. In fact, expanding universes satisfying this
assumption have $\dot{\rho}=-3H\left( P+\rho \right) <0$.

Problems may in principle arise with the better-known inflationary
attractor (for both $K=-1$ and $K=0$ universes), de Sitter space
\be
a(t)=a_0 \, \mbox{e}^{H\,t} \;,\;\;\;\;\;\;\;\;\; H=\mbox{const.} \;,
\ee
that corresponds to constant Hubble radius $H^{-1}$ and density and pressure
$ \rho=\Lambda/\kappa=-P $, where $\Lambda$ is the cosmological constant.

Being a fixed point in phase space, de Sitter space is forever removed from
Minkowski space, but there are good reasons  to believe that its total
energy is the same as for Minkowski space. By
considering a scalar 
field $\phi(t) $ nonminimally coupled to gravity, one
can obtain a {\em superinflationary} regime with $\dot{H}>0$; this is
impossible when the
scalar $\phi (t) $ couples minimally to gravity (Faraoni 2002). 
The existence of superinflationary solutions with nonminimally coupled
scalars was
demonstrated  both exactly and with numerical methods in
(Gunzig {\em et al.} 2001{\em a}, Rocha-Filho {\em et al.} 2000, 
Gunzig {\em et al.} 2001{\em b}). They find   
solutions emerging
from points arbitrarily close to Minkowski space, and therefore with the
same total
energy, that expand and are attracted by de Sitter space. These solutions have
constant energy, and hence {\em the energy $E_{dS}$ of the 
asymptotic de Sitter space must coincide with that of Minkowski space} 
\footnote{The homoclinic solutions mentioned above 
that emerge from initial points arbitrarily close to the Minkowski fixed
point in the infinite past, and return to it in the far future, are those
that escape the attraction basin of the de Sitter attractor considered
here. In this subsection  only heteroclinic solutions
connecting
different fixed points are considered.}.

Since the only quantity characteristic of the de Sitter metric is the Hubble
constant $H$, one might expect that the energy of de Sitter space $E_{dS}$
could depend on the value of $H$, and therefore, in principle,
$E_{dS}=E_{Minkowski}$
could be true only for special values of $H$. However this is not the
case, as it was shown (Gunzig {\em et al.} 2001{\em a}, 2001{\em b}) that
for {\em any} value $H_0$ of the de Sitter fixed point $\left( H_0,
\phi_0 \right)$, one can find solutions emerging from points arbitrarily
close to Minkowski space and converging to de Sitter space. Since the
total energy of these solutions is constant and initially $E_{Minkowski}$,
it must be $E_{dS}=E_{Minkowski}$ for any value of $H$.

If Minkowski space has zero energy, the result 
holds true for all de Sitter spaces, whatever the source of gravity may be (a 
scalar field coupled minimally or nonminimally, a cosmological constant, exotic
dark matter, supergravity fields, etc.), and also for all the solutions in the
attraction basin of the de Sitter fixed point. This result is compatible
with a recent paper 
(Kastor and Traschen 2002) introducing a non-negative energy 
for universes that are asymptotically de Sitter\footnote{We 
acknowledge a referee for pointing out this reference.}. This is best seen 
by considering the Schwarzschild-de Sitter metric
\be
ds^2=-\left( 1- \frac{2M}{r} -\frac{\Lambda r^2}{3} \right) dt^2
+ \left( 1-\frac{2M}{r} -\frac{\Lambda r^2}{3} \right)^{-1} dr^2 
+r^2 \left( d\theta^2 +\sin^2 \theta \, d\varphi^2 \right) \;,
\ee
which is asymptotically de Sitter. The Kastor-Traschen conserved charge, constructed 
from conformal asymptotic Killing vectors, is $
E_{KT}=M \, a(t) $.  In the limit $M\rightarrow 0$, in which de Sitter 
space is recovered, the conserved charge vanishes.

The previous considerations are particularly relevant because de Sitter spaces
are also attractors for anisotropic universes. It follows from the cosmic
no-hair theorems that inflation is a generic phenomenon (see Goldswirth \&
Piran (1992) for a review). A wide variety of initial conditions
including initial anisotropy leads to the same de Sitter-like expansion.
Then, the energy of a Bianchi model converging to a de Sitter solution
with zero
energy must also be constant, and equal to zero. This result is
compatible with pseudotensor-based claims of zero total energy for Bianchi
models (Radinschi 1999, Xulu 2000).

\section{Discussion and conclusions}
\setcounter{equation}{0}

At this point, one could argue that, since the energy of the universe is
constant and the zero level is arbitrary, one can {\em choose} as zero the
energy of Minkowski space. This is the only possible choice 
compatible with the argument presented in the Introduction. 
Since the FRW metric $g_{ab}(t)$ approaches the Minkowski metric
$\eta_{ab}$ (which has zero energy)  as $t \rightarrow +\infty$, {\em the
energy $E$ associated with $g_{ab}$ must also be identically zero at all
times}. 
This argument applies to any cosmological
metric, be it FRW or not, that has Minkowski space as its asymptotic limit
at large times or in the infinite past and has constant energy (provided
that the time parameter can be given  an unambiguous geometrical meaning,
as in the case of an FRW metric).

There are now different arguments pointing to the result of constant and
vanishing total energy  for open and critically open FRW spacetimes. The
arguments provided in this paper support previous pseudotensor-based
claims. 
The pseudotensorial methods as well as those in Cooperstock \& Israelit (1995) 
indicate 
a zero value for closed FRW universes. The idea that the 
universe has zero energy is embodied
in the Wheeler-de Witt equation of quantum cosmology, which corresponds to
the
eigenvalue problem for the Hamiltonian of a quantum universe with zero energy
eigenvalue (e.g. Kolb \& Turner 1994).

One might question the notion that an infinite spacetime filled with matter
should have zero energy. The instinctive answer might be that the energy is
infinite. However,  the open FRW 
universe dilutes its matter density to an infinite extent and is hence
asymptotic to 
Minkowski spacetime. We know that the latter has zero energy because the 
Schwarzschild spacetime with energy $m$ becomes the Minkowski spacetime when
$m=0$.

Even in an infinite spacetime such as Schwarzschild spacetime, there
is always an { \em effective} boundary in the sense that, even for a
Schwarzschild mass that expands forever to infinite dilution, there exists
a sphere that encloses all the matter
at any time $t$. This is what allows us to determine the non-vanishing 
global energy $m$ of the spacetime. By contrast, in FRW there is no such
boundary 
available at any $t$. Interestingly, this is the case for a finite closed 
($K = +1$) model as well. The vanishing of total energy in the case of a 
closed universe might be expected by analogy with the necessarily vanishing 
of total charge for a closed universe, i.e. every closed surface in a 
finite space
encloses a finite volume of space on either side. Thus the electric flux 
through the surface equals the total charge in the interior as well as to the 
same amount of total charge in the exterior with the necessarily opposite 
sign (Landau \& Lifshitz 1989).
There is the suggestion that it is the
property
of unboundedness that is the crucial element in rendering these universes, both
open and closed, with zero energy.

The results developed in this paper that the energy of the universe is 
constant and zero for open or critically open
FRW universes, and for Bianchi models evolving into de Sitter spacetimes,
should
not be regarded as merely technical. Indeed it is well to question why 
universes that are so different 
all have zero total energy.  One could speculate that this fact might be
related
to the problem of the origin of the universe. Indeed, since the universe is
by definition an isolated system, the zero energy result is compatible with 
the universe emerging from a ``system'' with zero energy, be it quantum vacuum
(Albrow 1973, Tryon 1973, Guth 1981), ``nothing'' (Vilenkin 1983), flat empty
space
(Prigogine {\em et al.} 1988, 1989, Gunzig {\em et al.} 2001{\em a}, 2001{\em
b}), or something else. In such a picture,
matter particles would have to be created at the expense of the gravitational
field
energy (e.g. Prigogine {\em et al. } 1988, 1989).
It seems inconceivable that the cosmos could emerge from any  physical system that
has nonvanishing total energy. This would require an exchange of energy between
the universe and a third system, making a cosmological spacetime an open system
from the thermodynamical point of view.

\section*{Acknowledgments}

We acknowledge a referee for helpful comments. 
V.F. was supported by the NATO Advanced Fellowship Programme through the
National Research Council of Italy (CNR); F.I.C. acknowledges partial
support from the Natural Sciences and Engineering Research Council of
Canada (NSERC). 

\clearpage          

{\small 
\begin{center} {\bf REFERENCES} \end{center}

\vskip1.5truecm

\noindent  Albrow, M.G. 1973, {\em Nature} 241, 56 \\\\
Amendola, L.,  Litterio, M.  \& Occhionero, F. 1990, {\em 
Int. J. Mod. Phys. A} 5, 3861\\\\
Banerjee, N. \& Sen, S. 1997, {\em Pramana J. Phys.} 49,
609\\\\
Buchbinder, I., Odintsov, S.D. \& Shapiro, I. 1992, 
{\em Effective Action in Quantum Gravity} (Bristol: IOP Publishing) \\\\
Callan, C.G. Jr., Coleman \& Jackiw, R. 1970, {\em Ann. Phys. (NY)} {\bf 59}, 42\\\\
Cooperstock, F.I. 1994, {\em Gen. Rel. Grav.} 26, 323\\\\
Cooperstock, F.I. \& Israelit, M.I. 1995, {\em Found.
Phys.}  25, 631\\\\
Faraoni, V. 2002, {\em Int. J. Mod. Phys. D} 11, 471\\\\
Foster, S. 1988, {\em Class. Quant. Grav.}  15, 3485\\\\
Goldswirth, D. \& Piran, T. 1992, {\em Phys. Rep.}
214, 223\\\\
Grib, A.A. \& Poberii, E.A. 1995, {\em Helv. Phys. Acta} {\bf 68}, 380\\\\
Grib, A.A. \& Rodrigues, W.A. 1995, {\em Gravit. Cosmol.} {\bf 1}, 273\\\\
Gunzig, E., Faraoni, V., Rocha Filho, T.M., Figueiredo, A. \& 
Brenig L. 2000, {\em Class. Quant. Grav.} 
17, 1783\\\\
Gunzig, E., Saa, A., Brenig, L., Faraoni, V.,  Rocha Filho, T.M. 
\& Figueiredo, A. 2001{\em a}, {\em Phys. Rev. D} 63, 067301\\\\
Guth, A.H. 1981, {\em Phys. Rev. D} 23, 347\\\\
Johri, V.B., Kalligas, D, Singh, G.P. \&  Everitt, C.W.F. 1995, {\em
Gen. Rel. Grav.} 27, 313\\\\
Kastor, D. \& Traschen, J. 2002, preprint hep-th/0206105\\\\
Kolb, E.W. \& Turner,  M.S. 1994, {\em The Early Universe}
(Reading, MA: Addison-Wesley)\\\\
Landau, L.D.  \&  Lifshitz, E.M.  1989, {\em The Classical Theory of Fields} 
(4th revised edition, Oxford: Pergamon Press)\\\\
Liddle, A.R.  \& Lyth, D. 2000, {\em Cosmological Inflation 
and Large Scale Structure} (Cambridge: Cambridge University Press)\\\\
Madsen, M.S. \&  Ellis, G.F.R. 1988, {\em MNRAS} 234, 67 \\\\
Madsen, M.S.,  Mimoso, J.P,  Butcher, J.A. \& Ellis, G.F.R. 1992, {\em Phys. Rev.
D} 46, 1399\\\\
Parker, L. \& Toms, D.J. 1985, {\em Phys. Rev. D} {\bf 32}, 1409\\\\
Prigogine, I., Geheniau, J., Gunzig, E.  \&  Nardone, P. 1988, 
{\em Proc. Natl. Acad. Sci. USA} 85,
7428\\\\
Prigogine, I., Geheniau, J., Gunzig, E.  \&  Nardone, P. 1989, 
{\em Gen. Rel. Grav.} 21, 767\\\\
Radinschi, I. 1999, {\em Acta Phys. Slov.} 49, 789 
(preprint gr-qc/0008034) \\\\
Rocha Filho, T.M., Figueiredo, A., Brenig, L., Gunzig, E. \&
Faraoni, V. 2000, {\em Int. J. Theor. Phys.} 39, 1933\\\\
Rosen, N. 1994, {\em Gen. Rel. Grav} 26, 319\\\\
Saa, A., Gunzig, E., Brenig, L., Faraoni, V.,  Rocha Filho, T.M. \&  Figueiredo,
A. 2001{\em b}, {\em Int. J. Theor. Phys.} 40, 2295 \\\\
Stabell, R. \& Refsdal, S.  1966, {\em Mon. Not. R. Astr.
Soc.} 132, 379\\\\
Sonego, S. \& Faraoni, V. 1993, {\em Class. Quant. Grav.} {\bf 10}, 1185\\\\
Tryon, E.P.  1973, {\em Nature} 246, 396\\\\
Vilenkin, A. 1983, {\em Phys. Rev. D} 27, 2848 \\\\
Wainwright, J. \&   Ellis,  G.F.R. 1997, {\em Dynamical
Systems in Cosmology} (Cambridge: Cambridge University Press)\\\\
Weinberg, S. 1972, {\em Gravitation and Cosmology} (New York: Wiley) \\\\
Xulu, S. 2000, {\em Int. J. Theor. Phys.} 39, 1153 (preprint
gr-qc/9910015)

}

\end{document}